\documentclass[a4paper,11pt]{article}
\pdfoutput=1 

\usepackage{jinstpub} 
\usepackage{subfigure}

\title{\boldmath 
Observation of thermal events on the plasma facing components of Wendelstein 7-X
}

\author[a,1]{A. Puig Sitjes \note{Corresponding author.}}
\author[b]{Y. Gao}
\author[a,e]{M. Jakubowski}
\author[a]{P. Drewelow}
\author[a]{H. Niemann}
\author[a,2]{A. Ali \note{Current affiliation: Max Planck Institute of Molecular Physiology, Germany}}
\author[c]{V. Moncada} 
\author[d]{F. Pisano} 
\author[c]{T.T. Ngo} 
\author[d]{B. Cannas} 
\author[e]{M. Sleczka} 
\author[a]{and W7-X Team}

\affiliation[a]{
Max-Planck-Institut f{\"u}r Plasmaphysik, Wendelsteinstra{\ss}e 1, 17491 Greifswald, Germany.
}
\affiliation[b]{
Forschungszentrum J{\"u}lich GmbH, J{\"u}lich 52425, Germany.
}
\affiliation[c]{
CEA, IRFM, F-13108 Saint Paul-lez-Durance, France.
}
\affiliation[d]{
University of Cagliari, Piazza d{\textquotesingle}Armi, 09126 Cagliari, Italy.
}

\affiliation[e]{
University of Szczecin, Szczecin, Poland.
}

\emailAdd{aleix.puig.sitjes@ipp.mpg.de}

\abstract{
Long pulse operation of present and future magnetic fusion devices requires sophisticated methods for protection of plasma facing components from overheating. Typically, thermographic systems are being used to fulfill this task. Steady state operation requires, however, autonomous operation of the system and fully automatic detection of abnormal events. At Wendelstein 7-X (W7-X), a large advanced stellarator, which aims at demonstrating the capabilities of the stellarator line as a future fusion power plant, significant efforts are being undertaken to develop a fully automatic system based on thermographic diagnostics. In October 2018, the first divertor-based experimental campaign has been finished. One of the goals of this operation phase (named OP1.2) was to study the capabilities of the island divertor concept using an uncooled test divertor made of fine-grain graphite tiles. Throughout this campaign, it was possible to test the infrared imaging diagnostic system, which will be used to protect the actively water-cooled plasma facing components (PFCs) during the steady-state operation in the next experimental campaign.

An overview of the most relevant thermal events on the PFCs that were detected in OP1.2 using this system are presented. This includes events that limited operation during the campaign, like baffle hot spots and divertor overloads, events that are potentially critical in steady state operation like leading edges, events caused by the ECRH and NBI heating systems (shine-through hot spots and fast particle losses) and other events which are a common source of false alarms like surface layers. The detected thermal events are now part of an important and extensive image database which will be used to further automate the system by means of computer vision and machine learning techniques in preparation for steady-state operation, when the system must be able to detect dangerous events and protect the machine in real-time.
}


\proceeding{3$^{\text{rd}}$ European Conference on Plasma Diagnostics\\
  6$^{\text{th}}$-9$^{\text{th}}$ of May 2019\\
  Lisbon, Portugal}

\begin{document}
\maketitle
\flushbottom

\section{Introduction} \label{sec:Introduction}

\begin{figure}
	\centering
	\includegraphics[width=0.8\textwidth]{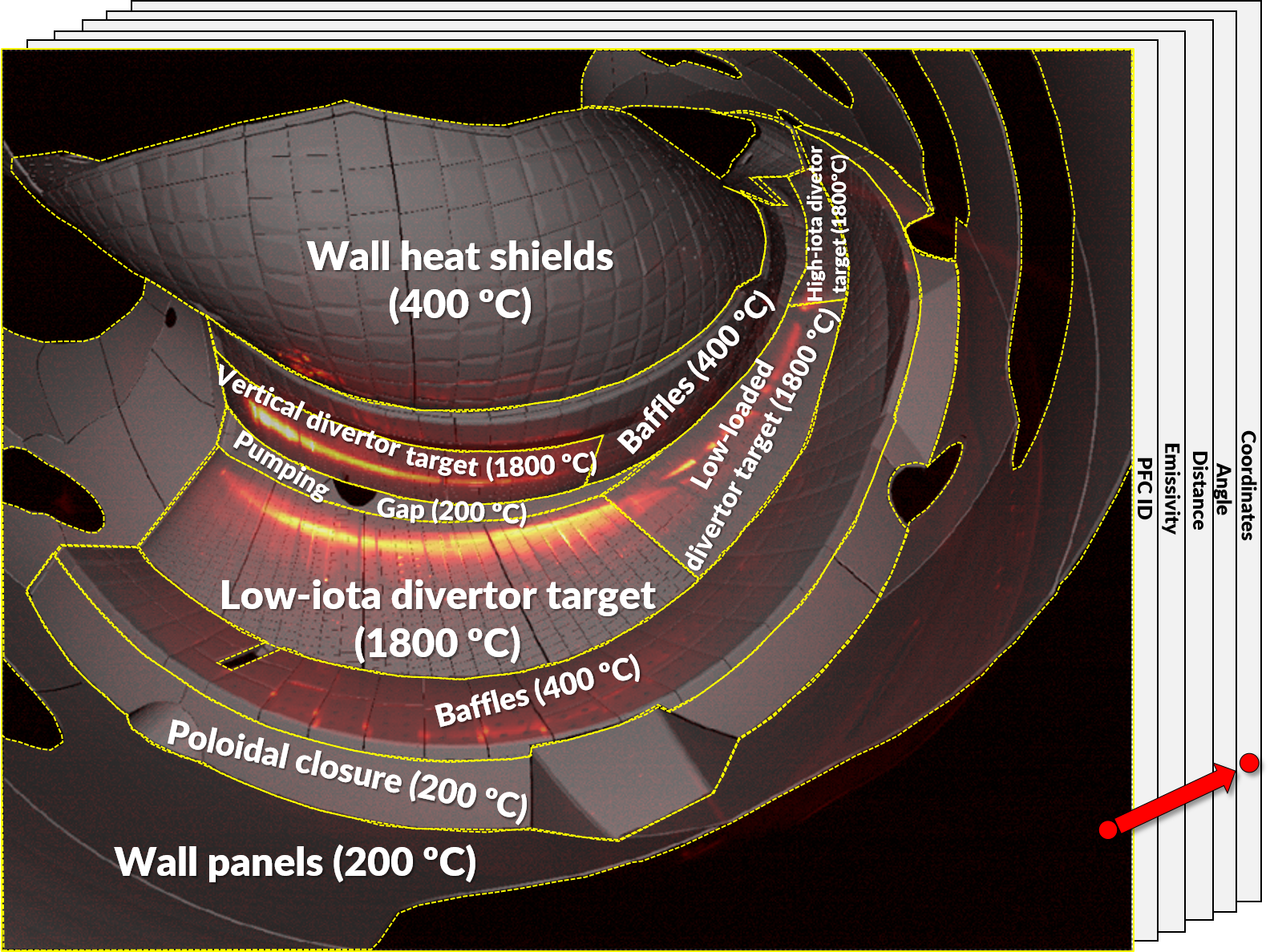}
	\hfill
	\caption{The PFCs as seen from the endoscope camera. The immersion tube cameras have the same view but with less distortion. The scene model and the OP1.2 operational limits are overlaid on top of the image.}
    \label{fig:pfc}
\end{figure}

The first divertor operational phase OP1.2 of Wendelstein 7-X (W7-X) concluded in October 2018 \cite{pedersen}. W7-X is composed of 5 symmetric modules equipped with 10 divertors, i.e., 5 upper and 5 lower units. During this operational phase inertially-cooled test divertor units (TDUs), made of fine-grain graphite and with the same geometry as the water-cooled divertor, were used. In addition, a thermography diagnostic was commissioned to provide real-time protection of plasma facing components (PFCs) \cite{jakubowski} in order to avert the danger of damaging the high-heat flux (HHF) divertor, able to withstand 10 MW/m\textsuperscript{2}. Apart from observing the divertors, the thermography diagnostic observes other PFCs such as the baffles, the inner wall heat shields as well as the outer wall and pumping gap panels. The baffles and heat shields are also made of fine grain graphite and are limited to a heat flux of 0.5 MW/m\textsuperscript{2} and a surface temperature of 400 $^{\circ}$C. The wall and pumping gap panels are made of stainless steel and are limited to 0.2 MW/m\textsuperscript{2} and 200 $^{\circ}$C (see Figure \ref{fig:pfc}).

The diagnostic in OP1.2 consisted in 10 infrared cameras installed in 9 immersion tubes and one prototype endoscope. The endoscope was an off-axis Cassegrain optical system with a pinhole aperture of 6 mm. The field of view was 115 x 60 deg with an optical resolution between 5 mm at the low-iota target and 20 mm at the high-iota target. The immersion tube cameras covered the spectral range from 8 to 10 {\textmu}m with 1024 x 768 pixels and the endoscope camera the spectral range from 2 to 5.7 {\textmu}m with 1280 x 1024 pixels, both running at 100 Hz.  The diagnostic was controlled by a software running on a distributed system of 10 acquisition and analysis workstations and one central workstation for visualization and control of the cameras. The software used a multi-threaded processing pipeline architecture to run algorithms for hot spot detection and classification on a GPU.

\section{Detection of thermal events} \label{sec:thermal_events}

In OP1.2 the diagnostic was used to protect the device during operation which required prompt detection of thermal events and checking their temperatures against the operational limits. The detected thermal events include strike-lines, leading edges, overload hot spots, shine-through hot spots due to the heating systems, fast particle losses and surface layers. The detection algorithm uses a scene model that provides pixel-wise information of the PFCs (see Figure \ref{fig:pfc}), which requires a precise geometric calibration of the cameras to fit the CAD model to the strongly distorted images \cite{pisano}. The surface temperature of the thermal event is computed taking into account for each pixel the target emissivity, its distance and the angle of the line of sight respect the normal to the surface. The world coordinates are then used to measure the physical size and the position of the event in order to evaluate its risk.

When a critical thermal event occurs in steady state operation in OP2, the reaction must be carried out in 1s in order to prevent damage to the device. The imaging system will have to detect the thermal events within 100 ms and send alarms to the interlock system which is designed to react within 50 ms. The temperature alarm threshold $T_{alarm}$ will be computed according to the reaction time required $t_r$, the maximum temperature allowed $T_{max}$ and the net measured heat-flux $q'=q-q_{cooling}$ (see \eqref{eq:alarms} where $C_m[s(kW)^2m^{-4}K^{-2})]$ is the thermal capacity of the component). A warning will be sent to change the scenario $t_r=1s$ before reaching the operational limit and an alarm will be sent to stop operation $t_r=150 ms$ in advance.

\begin{equation}
\label{eq:alarms}
T_{alarm}[K] = T_{max}[K]-q'[kWm^{-2}]\sqrt{\frac{t_r}{C_m}}
\end{equation}

\subsection{Hot spots} \label{sec:hot_spots}

\begin{figure}
	\hfill
	\subfigure[Program 20180823.037 (AEF31). Max. 802 $^{\circ}$C]{\includegraphics[width=0.46\textwidth]{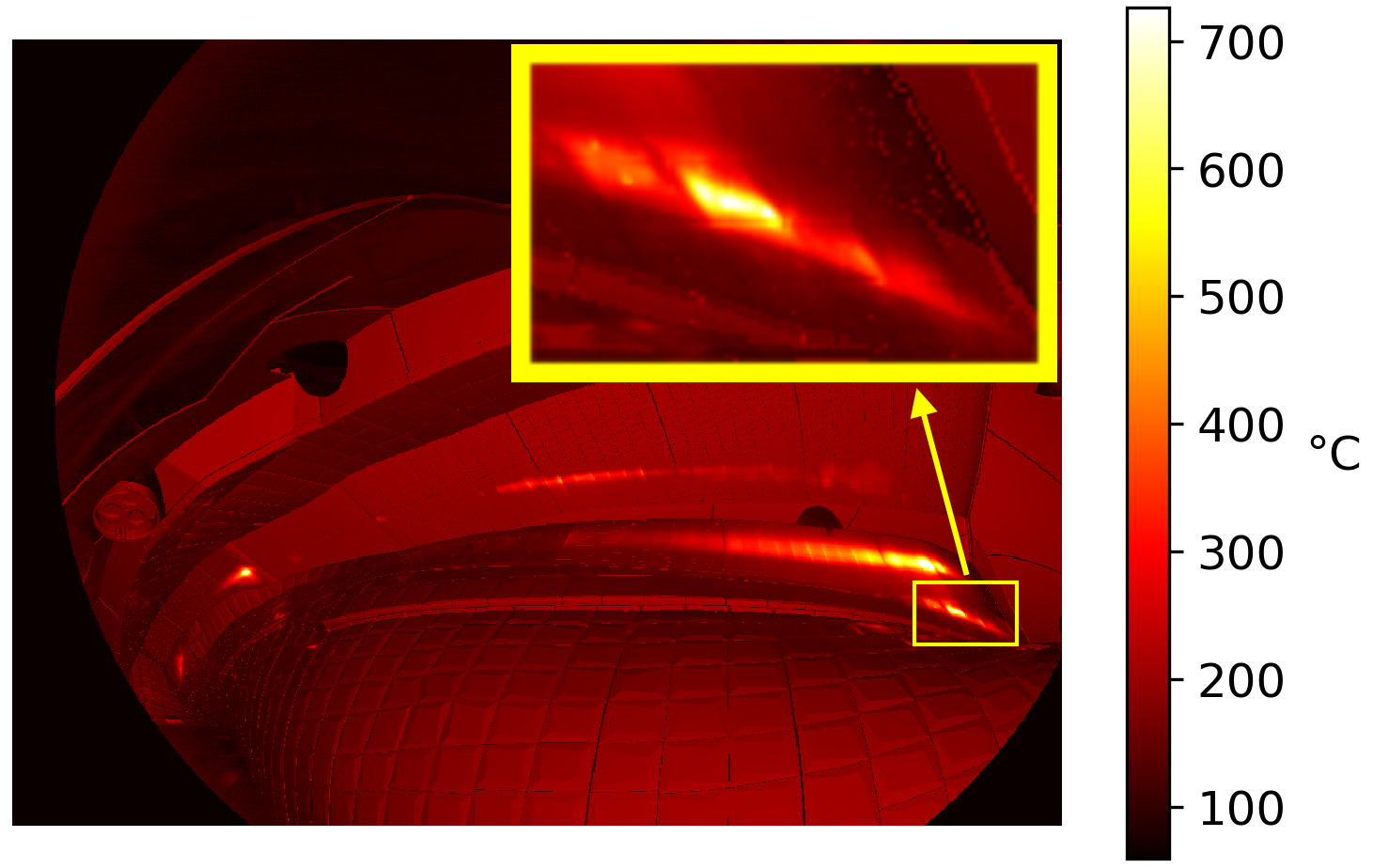}}
	\hfill	
	\subfigure[Program 20171114.053 (AEF20). Max. 456 $^{\circ}$C]{\includegraphics[width=0.46\textwidth]{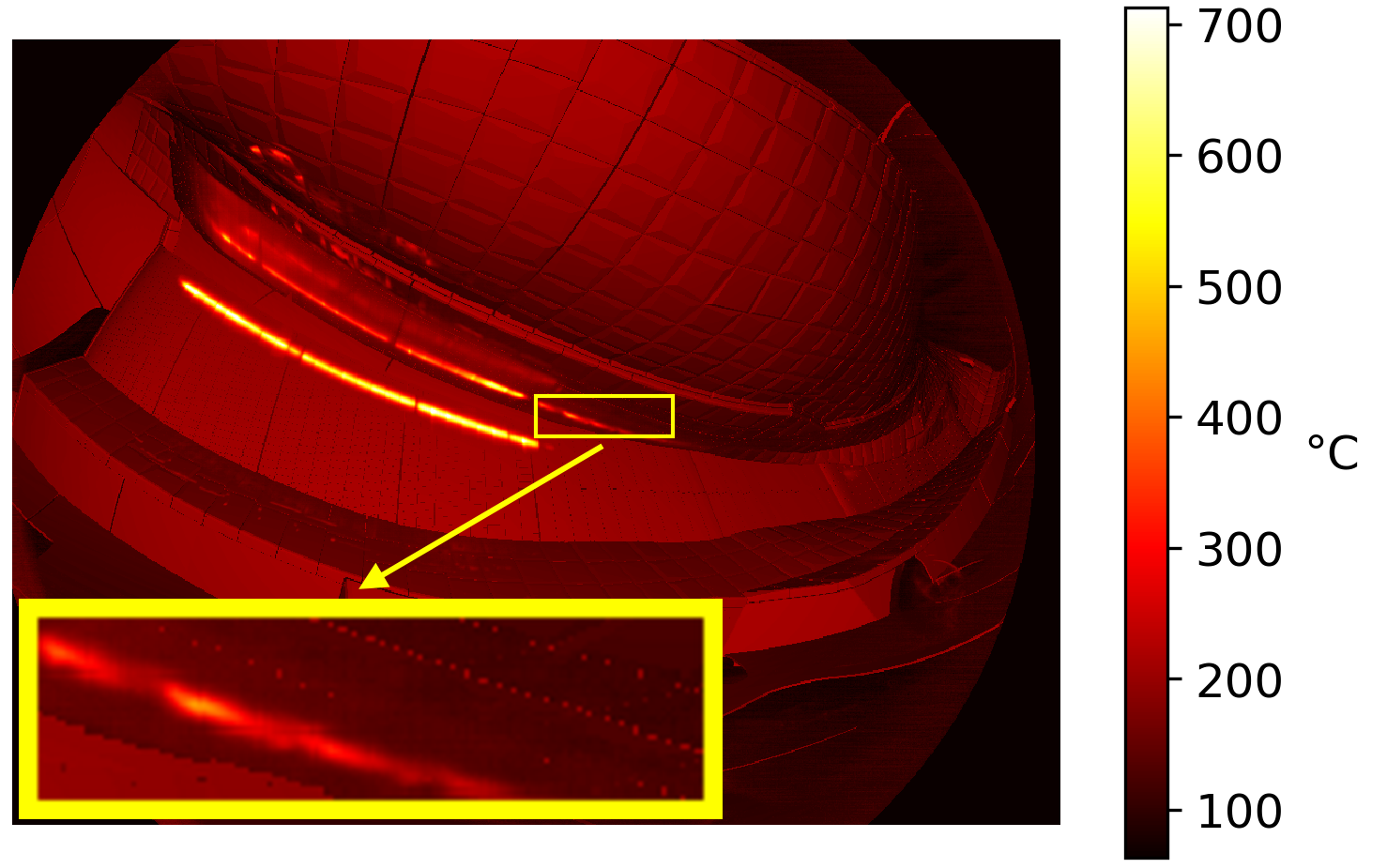}}
	\hfill
	\caption{(a) Hot spots were detected in the baffle above the vertical target in the high-mirror and standard configuration in all modules, but specially in module 3 in the high-mirror configuration. These hot spots were caused by an island at the edge, which is intersecting the baffle next to the divertor target. It is still unclear if this is due to a misalignment of the baffles or to the magnetic field structure caused by error fields. Some experiments were conducted to try to remove these heat loads by adjusting the control coil currents. By making the island bigger the heat loads were diverted to other targets in the same island chain. The heat load on the baffle was greatly reduced but the hot spot remained. (b) The strike-line in the vertical target extends towards the baffle in the high-mirror configuration due to a high current (>2.5 kA) in the control coils that enlarge the magnetic islands at the edge. This limited the maximum current that could be applied to the control coils in this configuration.}
    \label{fig:baffle_hot_spots}
\end{figure}

\begin{figure}
	\hfill
	\subfigure[Program 20180904.007 (AEF10). Max. 2356 $^{\circ}$C]{\includegraphics[width=0.46\textwidth]{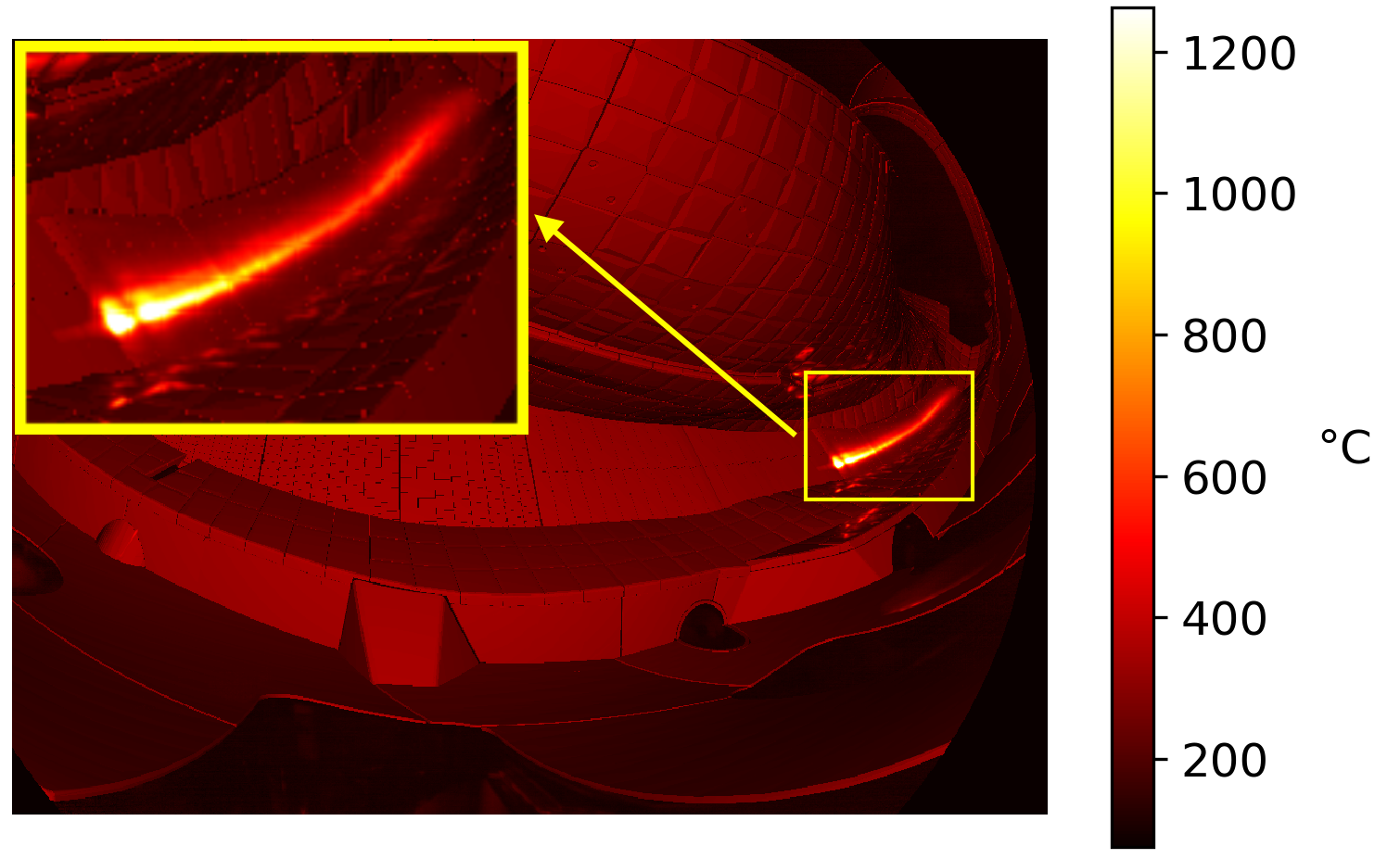}}
	\hfill
	\subfigure[Program 20180905.030 (AEF20). Max. 790 $^{\circ}$C]{\includegraphics[width=0.46\textwidth]{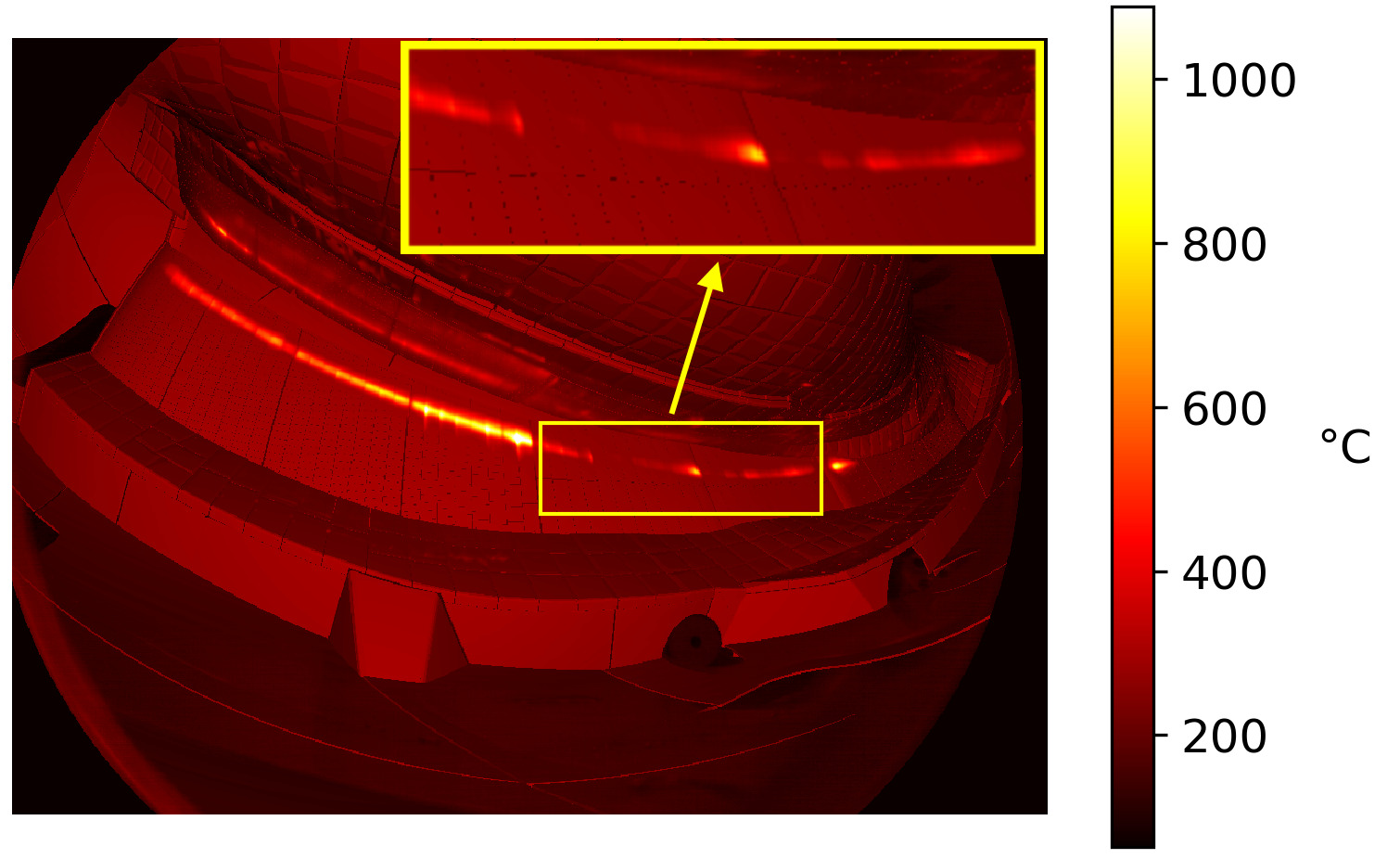}}
	\hfill
	\caption{(a) A hot spot is detected in the first finger of all high-iota divertor targets in the high-iota configuration. This hot spot is due to a step that exists between the low-loaded central target and the high-iota target. In this finger, the angle of incidence of the field lines is higher (with a maximum of 10.8$^{\circ}$ at the middle of the finger) and consequently the heat flux. This hot spot will limit the maximum heating power in steady-state operation at low densities for the high iota configuration if it is is not mitigated by detachment. (b) In the standard and high-mirror configurations, the strike-line extends towards the central part of the divertor, which has a baffle-like structure and it is not designed for high heat fluxes (it is limited to $1MW/m^2$, while $5.5 MW/m^2$ are detected in this experiment).}
    \label{fig:divertor_overload}
\end{figure}

Several thermal events were detected during the OP1.2 that limited operation and these will have to be addressed before the steady-state operation starts (see  Figure \ref{fig:baffle_hot_spots}). Other thermal events did not limit operation in OP1.2, but they could be potentially critical in high heat flux steady-state operation. In these cases, the detected temperatures did not exceed the operational limits of the inertially cooled TDU, but the heat loads would have exceeded the cooling capabilities of the components in OP2 (see Figure \ref{fig:divertor_overload}).

\subsection{Leading edges} \label{sec:leading_edges}

\begin{figure}
	\hfill
	\subfigure[IR image of experiment 20181017.038 (AEF10)]{\includegraphics[width=0.46\textwidth]{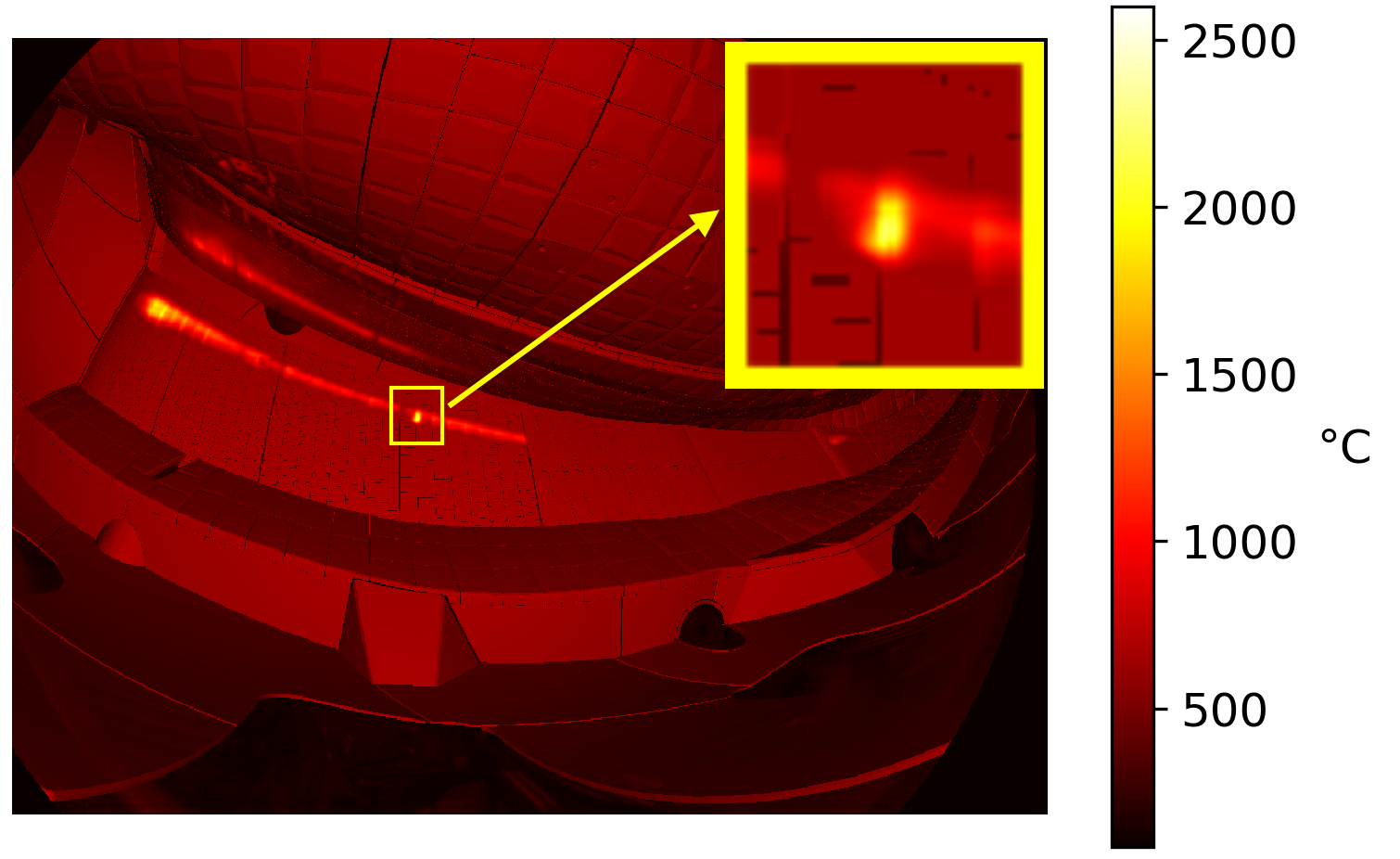}}
	\hfill
	\subfigure[The leading edge temperature over time.]{\includegraphics[width=0.4\textwidth]{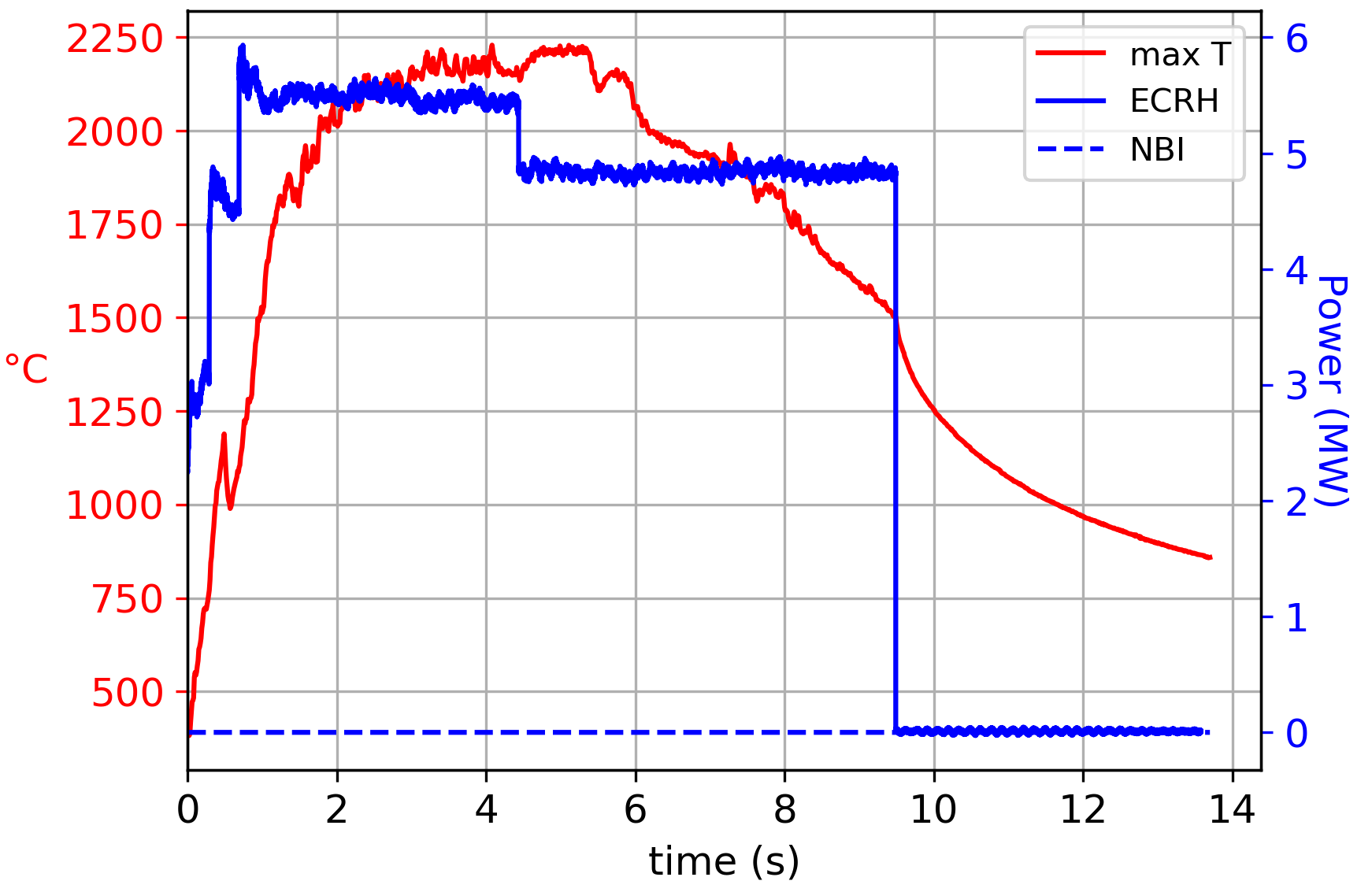}}
	\hfill
	\caption{A 10 mm long leading edge was overloaded reaching a maximum temperature of 2229 $^{\circ}$C.}
    \label{fig:leading_edges}
\end{figure}

In general, the divertor plates intersect the magnetic field lines at a shallow angle to effectively spread the power loads over larger areas, but misalignments of the divertor modules or fingers may result in prominent edges and, hence, higher heat loads that can exceed the operational limits of the tiles \cite{endler}. One of the goals of OP1.2 was to assess the effects of leading edges on the plasma performance, in preparation for the HHF divertor operation phase. Overloading of leading edges can lead to excessive release of carbon into the plasma and, as a result, cause a premature termination of the discharge, which can significantly hinder the steady state operation. Dedicated experiments, with increasing injected energy, were carried out in order to overload a well-known leading edge and evaluate its effect (see Figure \ref{fig:leading_edges}). Despite a significant carbon release, the plasma survived until the end of the discharge in all of the experiments, suggesting that the plasma edge effectively shields the plasma core. However, these impurities will accumulate in steady-state operation and overloading of leading edges must be avoided. Moving the strike-line position away from an overloaded leading edge can be achieved by means of the control coils, a special set of coils placed under the divertor modules which can change the topology of the magnetic islands at the edge. An advanced algorithm is currently under development to change in real-time the control coil currents and the strike-line position to avoid these overloads. 

\subsection{Shine-through hot spots and fast particle losses} \label{sec:shine_through}

\begin{figure}
	\hfill
	\subfigure[Program 20181009.041 (AEF10). Max. 734 $^{\circ}$C]{\includegraphics[width=0.46\textwidth]{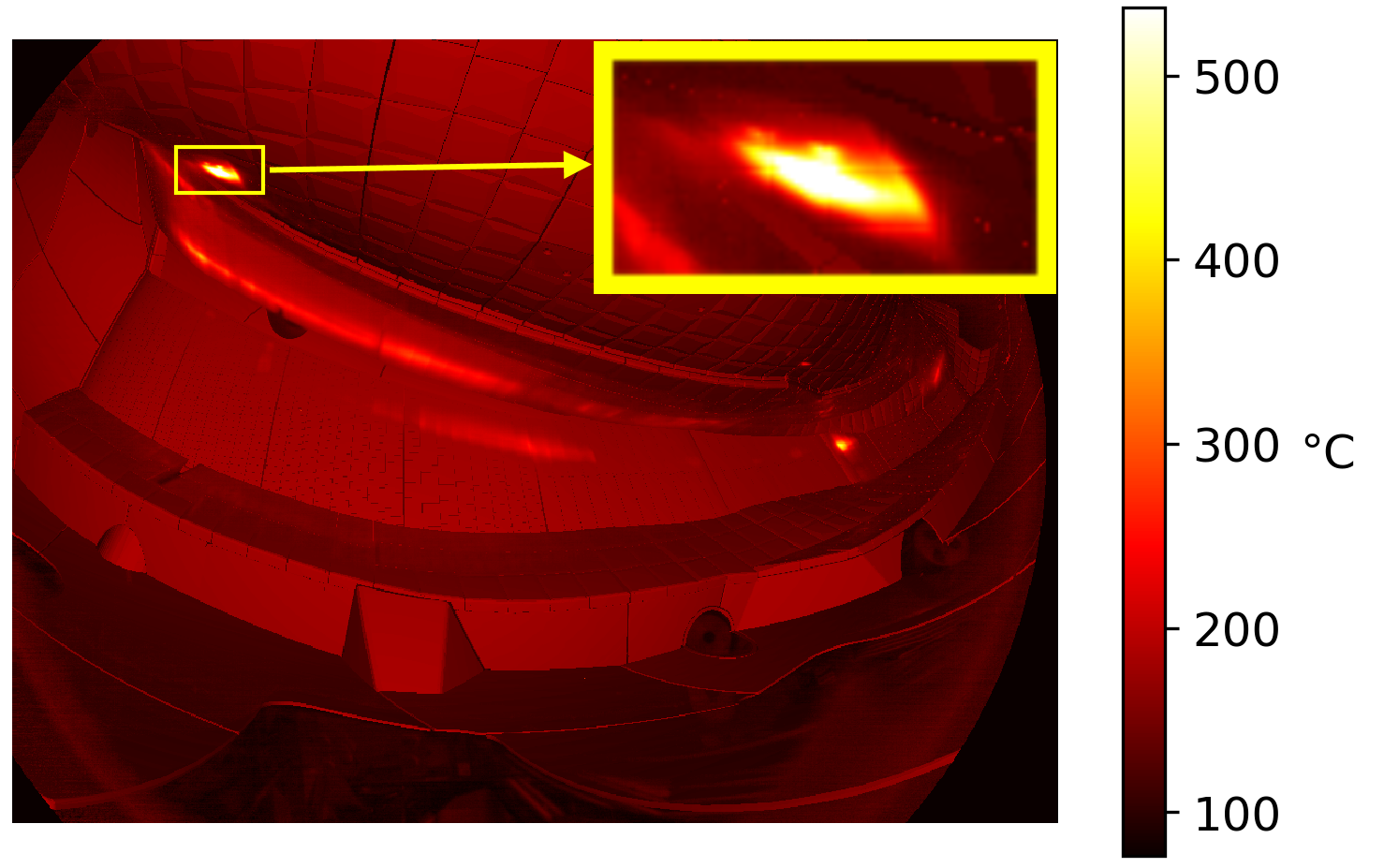}}
	\hfill
	\subfigure[Program 20180918.019 (AEF21). Max. 749 $^{\circ}$C]{\includegraphics[width=0.46\textwidth]{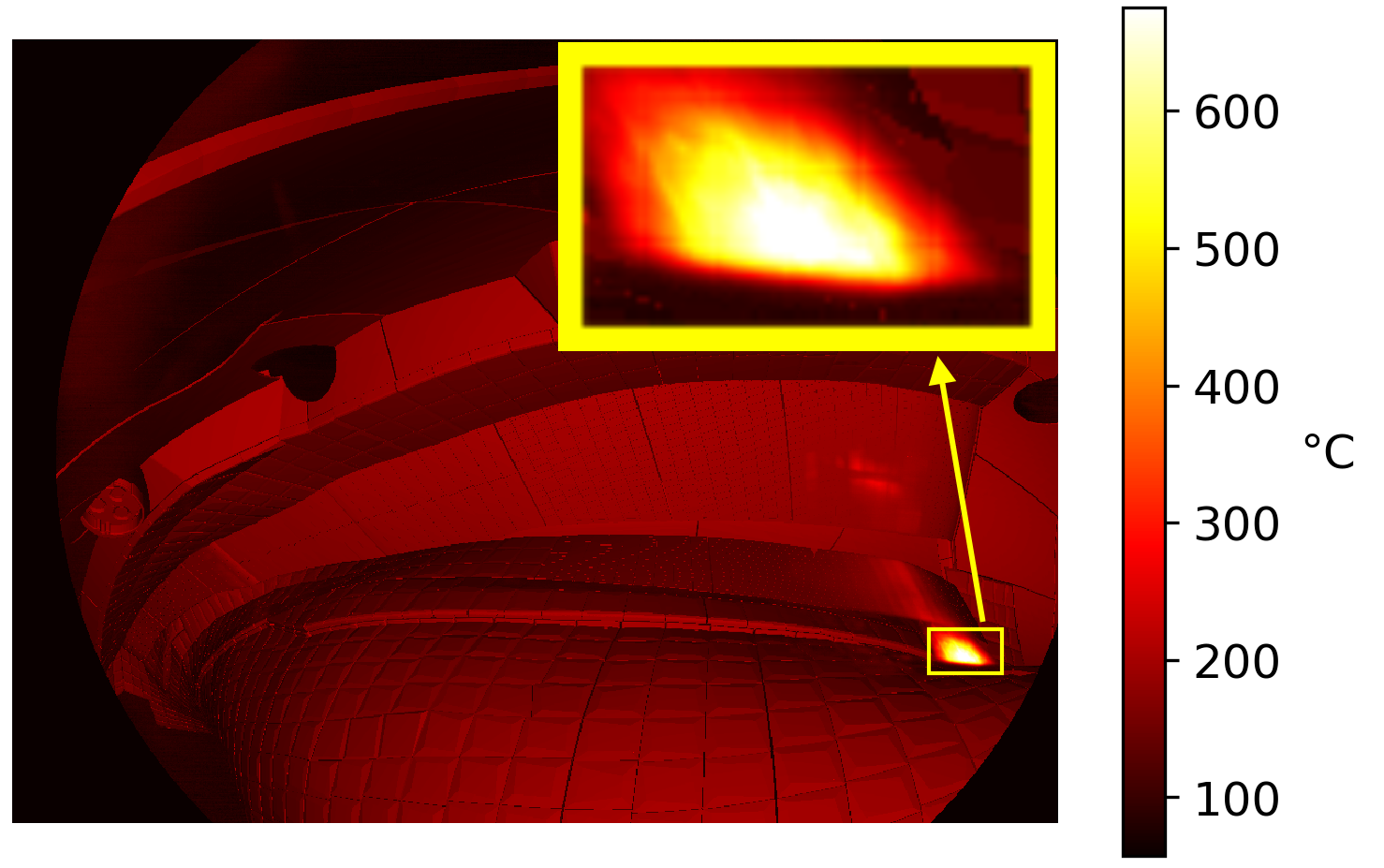}}
	\hfill
	\caption{(a) The ECRH causes an overload of the inner wall heat shields. (b) The NBI beam 7 hits the wall and a shine-through hot spot is detected the inner wall heat shields (failed experiment without plasma).}
    \label{fig:shine_through}
\end{figure}

Shine-through hot spots caused by the ECRH, which operated at a maximum power of 7.5 MW, and NBI heating systems, which operated at 1.8 MW for 5s, were also detected on the wall heat shields in OP1.2. The ECRH shine-through hot spots were observable with the infrared cameras placed at the opposite side of the inner wall. In OP1.2b the NBI heating system was used for the first time with one neutral injector and two different beams in operation (called source 7 and 8). Only the beam dump of source 7 was directly observable by an IR camera. Examples of these wall overloads can be seen in Figure \ref{fig:shine_through}.
 
\begin{figure}
	\hfill
	\subfigure[Program 20180919.037 (AEF31). Max. 397 $^{\circ}$C]{\includegraphics[width=0.46\textwidth]{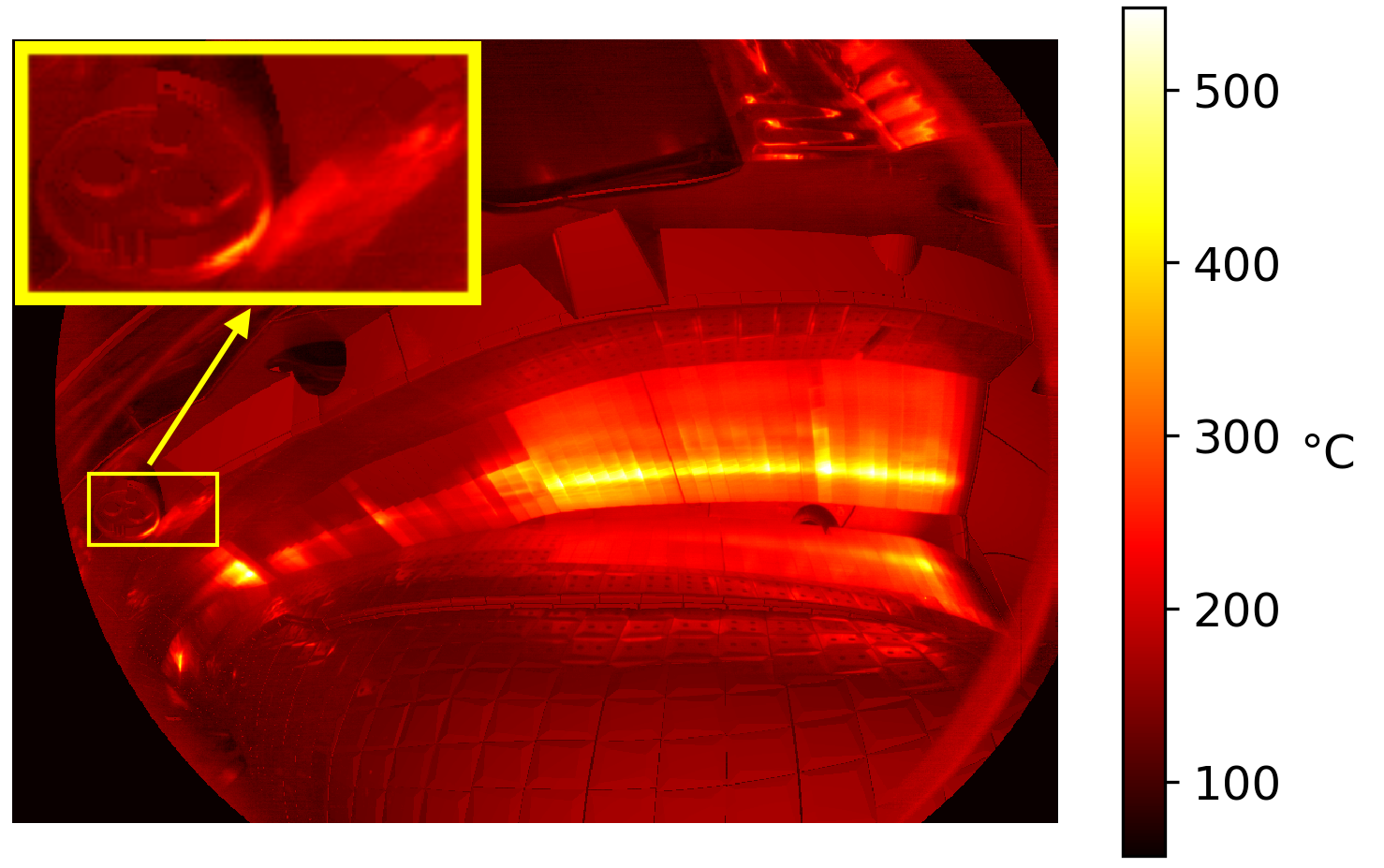}}
	\hfill
	\subfigure[Program 20180919.037 (AEF31). Max. 440 $^{\circ}$C]{\includegraphics[width=0.46\textwidth]{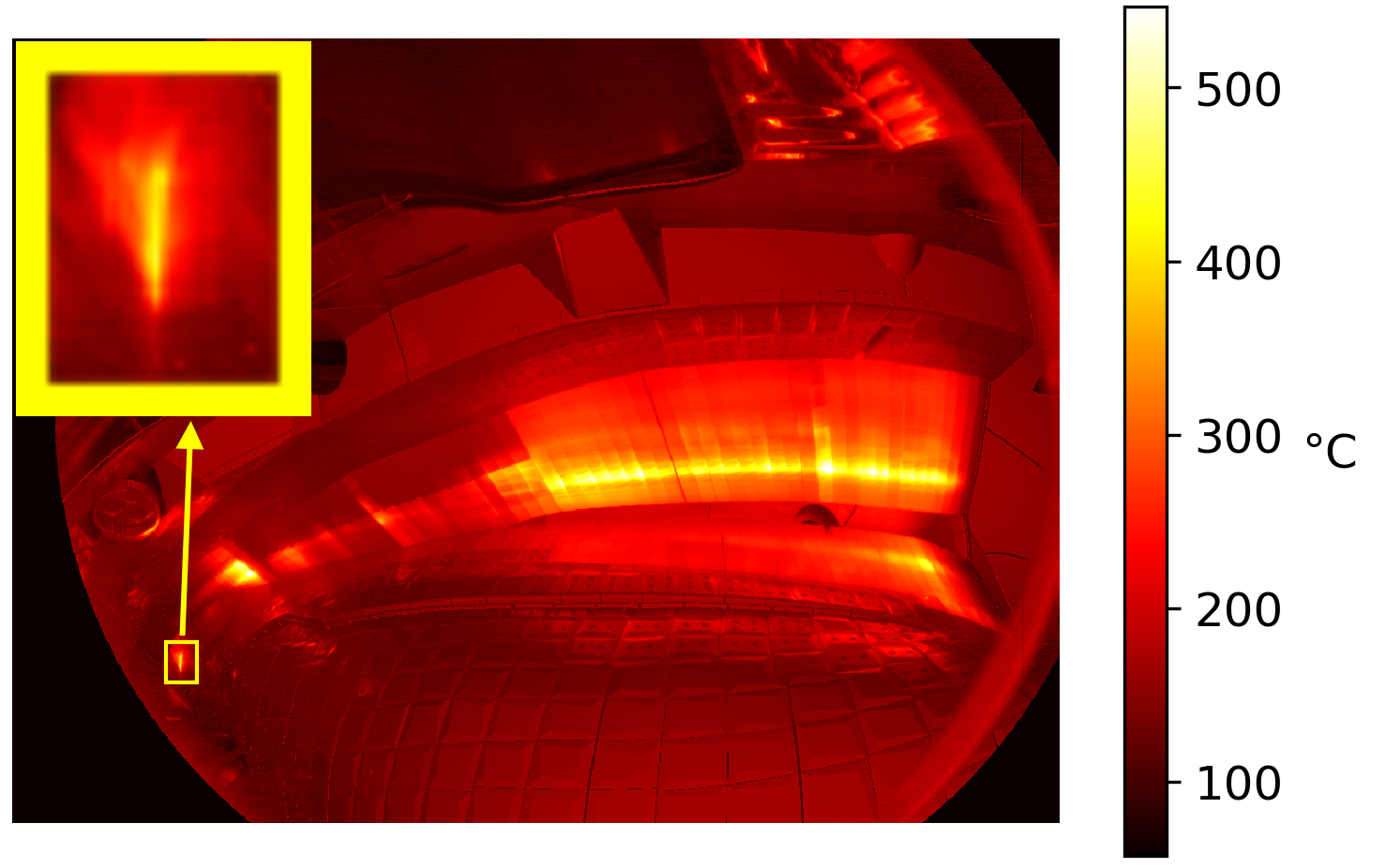}}
	\hfill
	\caption{(a) A hot spot due to fast ion losses from the NBI is detected on the protection collar of one of the immersion tubes. (b) A hot spot is detected on the baffle next to the high-iota divertor targets due to the NBI.}
    \label{fig:fast_particle_losses}
\end{figure}

According to simulations \cite{akaslompolo} hot spots caused by fast ion losses were expected at the walls, baffles and some diagnostic ports due to the NBI operation. The monitoring of these hot spots is of vital importance to guarantee a safe operation of the NBI (see Figure \ref{fig:fast_particle_losses}).

\subsection{Surface layers} \label{sec:surface_layers}

\begin{figure}
	\hfill
	\subfigure[Program 20180918.028 (AEF31)]{\includegraphics[width=0.5\textwidth]{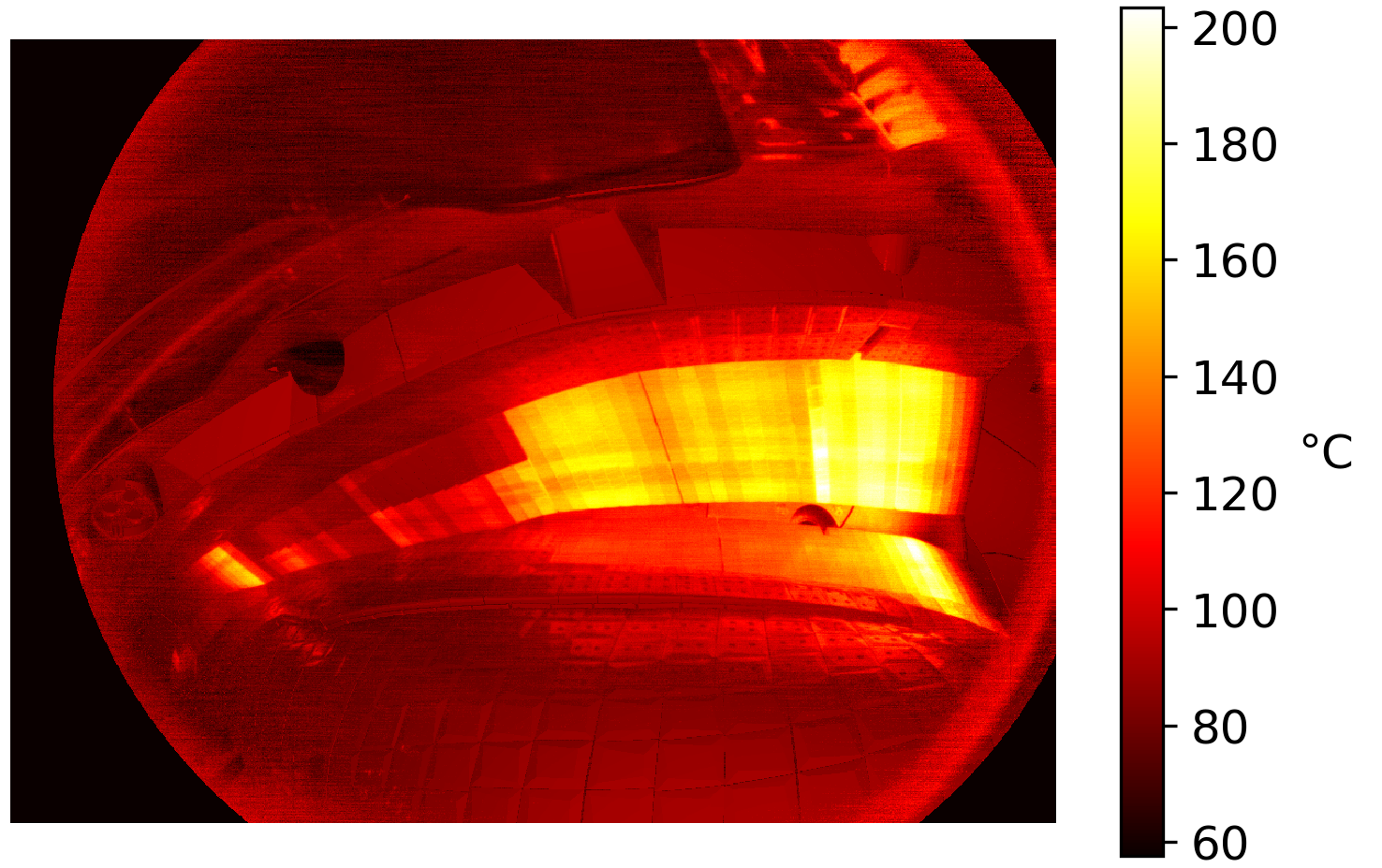}}
	\hfill
	\subfigure[Surface layers detail]{\includegraphics[width=0.4\textwidth]{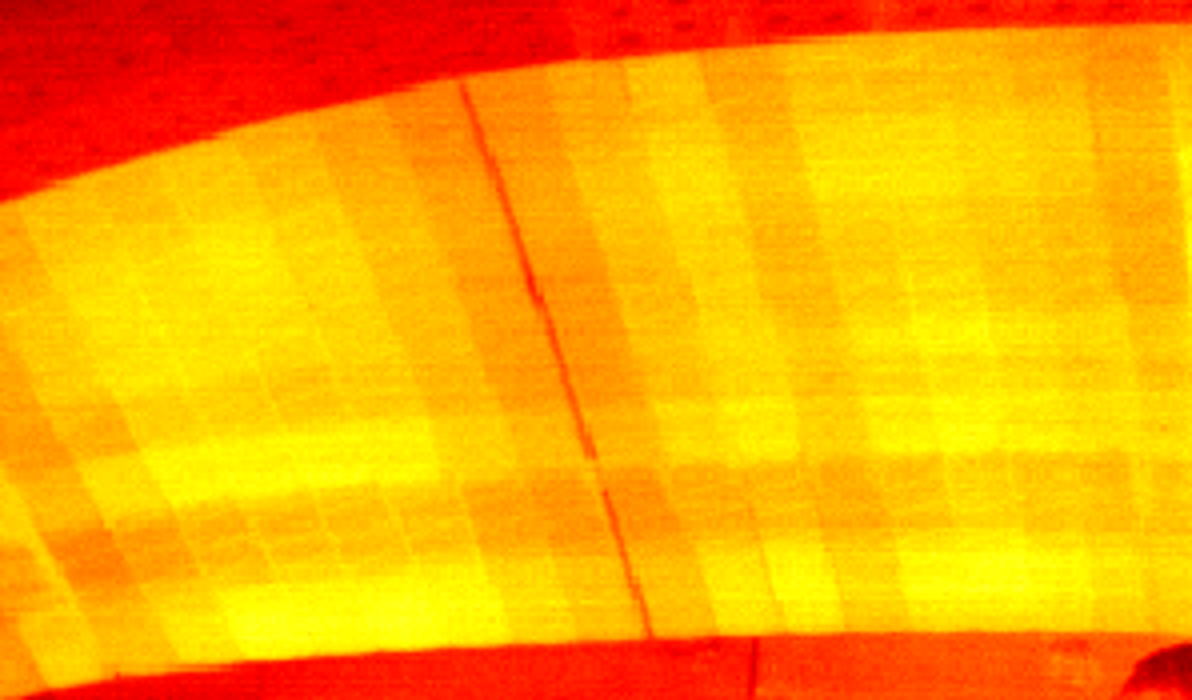}}
	\hfill
	\caption{The complex pattern of erosion and deposition layers on the divertor in thermal equilibrium.}
    \label{fig:surface_layers}
\end{figure}

Carbon erosion caused by the strike-line produces thin surface layers on the PFCs due to redeposition. They show as hot spots due to their low thermal capacity and their poor thermal connectivity to the underlying material. In W7-X, the magnetic configuration and the bootstrap current can change the strike-line position and, hence, the erosion-deposition patterns (see Figure \ref{fig:surface_layers}). This means that the imaging system has to detect the surface layers in real-time in order to avoid false alarms \cite{ali}.

%
%

\section{Conclusions} \label{sec:conclusions}

During OP1.2 significant thermal events were detected, some of which already exceeded the operational limits of the TDU, while others are potentially critical in future long plasmas. In steady-state, a real-time heat-flux computation, not yet available, will be required to send the alarm in order to react in time before reaching the temperature limit. The scene model concept, which provides a pixel-wise correspondence of the image with the PFCs limits, proved to be a better approach when hot spots can occur anywhere in the field of view, compared to region-based algorithms. However, the endoscope strong lens distortion was very difficult to model and there was a slight misalignment of the CAD in the high-iota region. Furthermore, the resolution of the optical system was insufficient to discern overloads from leading edges in that region of the divertor. To avoid leading edges, advanced real-time strike-line control algorithms will be necessary. Surface layers were also a challenging problem and improved algorithms have to be developed for their detection. All these events are now part of an extensive image database which will be used as a training dataset to further automate the protection system with computer vision techniques. The goal is to achieve a fully automated plasma control system for the steady-state operation, when the system must protect the water-coolded PFCs by changing the scenario and sending alarms to interlock system.

\acknowledgments

This work has been carried out within the framework of the EUROfusion Consortium and has received funding from the Euratom research and training programme 2014-2018 and 2019-2020 under grant agreement No 633053. The views and opinions expressed herein do not necessarily reflect those of the European Commission.


\begin{thebibliography}{99}


\bibitem{pedersen} T.S. Pedersen et al., \emph{First results from divertor operation in Wendelstein 7-X,  Plasma Physics and Controlled Fusion, Proc. of the 45th EPS Conference on Plasma Physics}, {\bf 61} (2018) 1.

\bibitem{jakubowski} M. Jakubowski et al., \emph{Infrared Imaging Systems for wall protection in the W7-X stellarator, Review of Scientific Instruments}, {\bf 89} (2018) 10E116.



\bibitem{pisano} F. Pisano et al., \emph{Towards a new image processing system at Wendelstein 7-X: From spatial calibration to characterization of thermal events, Review of Scientific Instruments}, {\bf 89} (2018) 123503.

\bibitem{endler} M. Endler et al., \emph{Managing leading edges during assembly of the Wendelstein 7-X divertor, Plasma Physics and Controlled Fusion}, {\bf 61} (2018) 2.

\bibitem{akaslompolo} S. {\"A}k{\"a}slompolo et al., \emph{Modelling of NBI ion wall loads in the W7-X stellarator, Nuclear Fusion}, {\bf 58} (2018) 8.

\bibitem{ali} A. Ali et al., \emph{Initial Results from the Hotspot Detection Scheme for Protection of Plasma Facing Components in Wendelstein 7-X, Nuclear Materials and Energy}, {\bf 19} (2019) 335\textendash339.

\end{thebibliography}
\end{document}